# Direct observation of monoclinic polar nanoregions in the relaxor ferroelectric Pb(Yb$_{1/2}$Nb$_{1/2}$)O$_3$–PbTiO$_3$

Hiroshi Nakajima[1], Satoshi Hiroi[2], Hirofumi Tsukasaki[1], Charlotte Cochard[3], Florence Porcher[4], Pierre-Eymeric Janolin[5], and Shigeo Mori[1, *]

[1]*Department of Materials Science, Osaka Metropolitan University, Sakai, Osaka 599-8531, Japan*

[2]*Diffraction and Scattering Division, Center for Synchrotron Radiation Research, Japan Synchrotron Radiation Research Institute, 1-1-1 Kouto, Sayo, Hyogo 679-5198, Japan*

[3]*School of Science and Engineering, University of Dundee, Nethergate, Dundee, DD1 4HN, United Kingdom*

[4]*Laboratoire Leon Brillouin, UMR12 CEA-CNRS, Bât. 563 CEA Saclay, 91191, Gif sur Yvette Cedex, France*

[5]*Université Paris-Saclay, CentraleSupélec, CNRS, laboratoire SPMS, 91190 Gif-sur-Yvette, France*

Address correspondence to E-mail: moris@omu.ac.jp

Relaxor ferroelectrics are applied in electronic devices such as actuators and sonars. Morphotrophic phase boundaries (MPBs) with monoclinic structures are known for their high piezoelectricity and electromechanical coupling factors in solid solutions of PbTiO$_3$ and relaxor ferroelectrics (Pb(Mg$_{1/3}$, Nb$_{2/3}$)O$_3$ or Pb(Zn$_{1/3}$, Nb$_{2/3}$)O$_3$). However, the monoclinic structures related to polar nanosize domains (polar nanoregions) exhibiting the relaxor properties of dielectric dispersion have not been reported. Using transmission electron microscopy and synchrotron x-ray scattering, we present the first observations of coexisting monoclinic structures and polar nanoregions near the MPB in Pb(Yb$_{1/2}$Nb$_{1/2}$)O$_3$–PbTiO$_3$. The polar nanoregions in this material are randomly shaped, unlike the ferroelectric nanodomains of the canonical relaxor Pb(Mg$_{1/3}$, Nb$_{2/3}$)O$_3$–PbTiO$_3$. Furthermore, *in situ* observations reveal that the monoclinic polar nanoregions grow as the temperature decreases. A pair-distribution function analysis reveals a mixture of monoclinic *Pm* and *Cm* structures in the polar nanoregions without the rhombohedral structure of other Pb-based relaxor solid solutions. Owing to the peculiar nature of the coexistence of the relaxor property (polar nanoregions) and high piezoelectricity (monoclinic structure), this material is expected as a new platform for understanding relaxor ferroelectricity.

**Introduction**

Revealing inhomogeneous polar nanostructures is a challenging task in materials science. However, such complex polar structures must be characterized because they enhance the functional properties of materials. Relaxor-ferroelectric solid solutions are a special class of materials exhibiting high piezoelectricity near the morphotropic phase boundary (MPB) separating the two phases [1,2]. These materials have a significant impact on piezoelectric devices because their piezoelectric coefficients exceed those of commercially-used Pb(Zr, Ti)$O_3$ by more than four times. Their properties can be understood through ferroelectric nanosized domains (*i.e.* polar nanoregions), which were postulated to explain the nonlinear temperature dependence of the index of refraction [3,4]. Recent studies have suggested that polar nanoregions enhance the piezoelectricity in relaxor-ferroelectric solid solutions of Pb(Mg$_{1/3}$, Nb$_{2/3}$)$O_3$–$x$PbTi$O_3$ and Pb(Zn$_{1/3}$, Nb$_{2/3}$)$O_3$–$x$PbTi$O_3$ [5]. However, polar nanoregions that are genuinely related to the relaxor properties and piezoelectricity enhancement remain unexplored because the relaxor characteristics of dielectric dispersion are lost as the PbTi$O_3$ content increases. Therefore, we focus on a sold solution that simultaneously induce the relaxor property and piezoelectricity due to PbTi$O_3$ substitution.

Pb(Yb$_{1/2}$Nb$_{1/2}$)$O_3$–$x$PbTi$O_3$ (PYN–$x$PT) is a solid solution of an antiferroelectric and a ferroelectric, and the relaxor and ferroelectric phases are separated by an MPB ($x$ = 0.5). The piezoelectric coefficient of PYN–$x$PT is high, being 510 pC/N for polycrystals and 2,500 pC/N for single crystals. The electromechanical coupling factor, permittivity, and remanent polarization of PYN–$x$PT are approximately 55%, 2,000 at room temperature (16,000 at the Curie temperature), and 30 μC/cm$^2$, respectively [6,7], comparable to those of typical ferroelectrics Pb(Mg$_{1/3}$, Nb$_{2/3}$)$O_3$ and Pb(Zr, Ti)$O_3$ used in actuators and sensors. A well-known feature of PYN–$x$PT is high-temperature stability with a Curie temperature of approximately 350°C. Accordingly, PYN–$x$PT is suitable for applications subjected to large strains and high operating temperatures, such as multilayer actuators [8,9].

PYN–$x$PT exhibits antiferroelectricity in the range $0 \leq x \leq 0.1$, relaxor ferroelectricity in the range $0.1 \leq x \leq 0.5$, and ferroelectricity in the range $0.5 \leq x$ [10,11]. Notably, both the high piezoelectricity and relaxor properties of dielectric dispersion are observed in the same relaxor phase. This property is unusual because the two phenomena are exclusive in other Pb-based solid solutions. For instance, increasing the PbTiO$_3$ content in $(1-x)$Pb(Mg$_{1/3}$, Nb$_{2/3}$)O$_3$–$x$PbTiO$_3$ suppresses the relaxor character while increasing the piezoelectricity [12]. Despite these interesting properties, the mechanism underlying the high piezoelectric coefficients and the nature of the microscopic polar nanoregions in the relaxor phase remain elusive. How the $x$ content influences the local structural change is also important for understanding this material.

Through transmission electron microscopy (TEM) imaging and x-ray pair-distribution function (PDF) analysis, we show that monoclinic polar nanoregions are connected to the relaxor property and high piezoelectricity of PYN–$x$PT. The low-symmetry polar nanoregions related to the relaxor property differ from those of typical relaxor-ferroelectric-based solid solutions. We also show that the size of these regions increases with decreasing temperature, leading to dielectric dispersion.

**Methods**

Dark-field images were obtained using a transmission electron microscope (JEM-2100F, JEOL Co. Ltd.) with an acceleration voltage of 200 kV. The images and diffraction patterns were recorded with a complementary metal–oxide semiconductor camera (OneView, Gatan Inc.). Heating was observed *in situ* using a heating holder (Gatan 648). The specimen was maintained for 10 min at each temperature (specified in the video of the heating experiment). Atomic-resolution high-angle annular dark-field (HAADF) scanning TEM (STEM), and bright-field STEM were conducted on a microscope equipped with a spherical-aberration corrector at an acceleration voltage of 200 kV (JEM-ARM200CF, JEOL Co. Ltd.). The probe semi-angle, current, and angular-detection range of the HAADF detector were 18.6 mrad, 9 pA, and 50–150 mrad, respectively. The observed specimens were fabricated via mechanosynthesis [11]. The PbO, YbNbO$_4$, and TiO$_2$ compounds were mixed in a ball-mill for 9 h and heated at approximately 950–1,050°C for 4 h. The specimens were thinned by argon ion milling and carbon-coated to prevent electron charging.

The dielectric constant $\varepsilon^* = \varepsilon' + i\,\varepsilon''$ was determined from impedance measurements on ceramics sputtered with gold electrodes. For these measurements, HP4294A and Agilent 4192 impedance analyzers were operated from 80 to 800 K in the frequency range 1 kHz to 1 MHz with a peak-to-peak bias of 400 mV. To differentiate the relaxor and ferroelectric behaviors, we measured the derivative of the inverse permittivity, $\xi = \frac{\partial 1/\varepsilon}{\partial T}$. This expression easily distinguishes a Curie–Weiss temperature dependence ($\xi$ independent of temperature) from relaxor behavior (linear temperature dependence of $\xi$), where the gradient is the exponent in the modified Curie–Weiss model). The temperature evolutions of some diffraction peaks were followed by x-ray diffraction on a high-accuracy Bragg–Brentano diffractometer at the Cu $K_\alpha$ wavelength issued from an 18-kW rotating anode. The peaks were fitted to pseudo-Voigt profiles.

The total x-ray scattering measurements were collected on the high-energy x-ray diffraction beamline BL04B2 at the SPring-8 facility. An incident x-ray beam of energy 113 keV was monochromated by the Si 333 reflection of a bent monochromator. The angular range of the measurements was 0.3–25°, giving a maximum momentum transfer $Q_{max}$ of approximately 25 Å$^{-1}$. The coherent intensity $I(Q)$ was obtained by subtracting the background, absorption, and polarization effects from the experimental scattering intensity. The structure factor $S(Q)$ was obtained by normalizing the experimental coherent scattering intensity based on the Faber–Ziman formula [13]. For calculating the experimental PDF $G(r)$, we Fourier-transformed $S(Q)$ as follows:

$$G(r) = \int_0^{Q_{max}} Q(S(Q-1)\sin(Qr)\,dQ. \quad (1)$$

The specimens were ground in a mortar and then packed into quartz capillaries. To avoid the preferred orientation effect, the capillaries were rotated during the measurements. Structural models were constructed by fitting to the experimental PDF [14]. The lattice parameters, atomic coordination, and temperature factors were refined using the PDF data.

## Results

We first characterize the structural change in the compound and then examine polar nanoregions in the relaxor phase. Figure 1 presents the electron diffraction patterns of PYN–$x$PT with different compositions. Based on the electrical measurements, the patterns were classified into three phases: antiferroelectric, relaxor ferroelectric, and ferroelectric [11]. In the antiferroelectric phase ($x = 0$ and 0.05), superlattice reflections were observed in all three directions. Along the [110] axis, the observed {½ ½ ½}-type superlattice reflections were attributed to cation ordering of Yb and Nb [15]. Along the [001] axis, superlattice reflections originating from antiferroelectric displacements were observed. The displacement vector described as (0.36, 0.36, 0) ran along the [110] and [1$\bar{1}$0] directions, indicating the presence of a grain boundary. The two directions of superlattice reflections in selected-area diffraction are presented in Supplementary Figs. 1 and 2. During the relaxor phase ($x = 0.2$ and 0.4), the antiferroelectric superlattice reflections along the [001] direction disappeared but the {½ ½ ½}-type reflections of cation ordering in the [110] patterns remained. In the ferroelectric phase ($x = 0.6$), both types of superlattice reflections disappeared and the (110) reflection was split.

The local structural changes in real space were observed through dark-field imaging and STEM. Figure 2 shows atomic-resolution STEM images of pure PYN along the [001] direction. Antiferroelectric displacements appear as modulations of the $(110)_{pc}$ lattice planes in the pseudocubic (PC) structure. The fast Fourier transform (FFT) shows that this modulation is related to the $(0.36\ 0.36\ 0)_{pc}$ superlattice reflection observed in the electron diffraction pattern. Specifically, the antiferroelectric displacements can be described as a motif with four up and four down dipoles pointing along the [±1 ±1

0]$_{pc}$ direction: ↑↑↓↑↓↓↑↓. Such a modulation of the antiferroelectric structure (Fig. 2c, inset) was observed in previous x-ray diffraction patterns [16,17]. The satellite peak position (0.36, 0.36, 0)$_{pc}$ of TEM is comparable with the peak (3/8, 3/8, 0)$_{pc}$ predicted by structural models. The displacements are perpendicular to the satellite peak direction of FFT, demonstrating a transverse modulation wave. In the HAADF–STEM image, the displacement of the arrow-marked Pb atoms is 0.27 Å, close to that of the x-ray structural analysis [16]. The STEM images in Fig. 2 reveal the complex and antiferroelectric structure of Pb(Yb$_{1/2}$Nb$_{1/2}$)O$_3$. It should be noted that electron irradiation can remove these antiferroelectric displacements (see Supplementary Fig. 3), converting the specimen into a simple perovskite structure. Because the specimen is insulating, the charging due to electron irradiation causes local electric fields [18,19]. These local electric fields should destabilize the local electric polarization of Pb atoms, which caused the transition from the antiferroelectric to paraelectric phases.

The effect of the {½ ½ ½} superlattice reflections on microstructure was also investigated through dark-field imaging. This study was performed on PYN–0.2PT, which is eminently suitable for ordering analysis owing to its low ferroelectric transition temperature ($T_c$ = 343 K); accordingly, it exhibits small spontaneous polarization at room temperature [10]. Figure 3 shows a dark-field image of the ½ ½ ½ reflection, demonstrating the presence of antiphase boundaries. In pure PYN, the Yb and Nb cations are ordered along the [111] axis [15], facilitating a phase shift of cation ordering and the subsequent creation of antiphase boundaries. Similar antiphase boundaries have been reported in lead perovskite Pb(Mg$_{1/3}$Nb$_{2/3}$)O$_3$ [20], which has an Mg:Nb ratio of 1:2. Accordingly, the Nb atoms should be segregated by {½ ½ ½} ordering [21]. Conversely, in Pb(Yb$_{1/2}$Nb$_{1/2}$)O$_3$ with an Yb:Nb ratio of 1:1, ordering is easily induced with no energy loss of segregation. The cation ordering comprises the sequence of Yb and Nb planes along the [111]$_{pc}$ direction (see the cation-ordering structure in Supplementary Figure 4). The intensities of the {½ ½ ½} spots decrease with increasing the $x$ content as shown in Fig. 1 and Supplementary Fig. 4(c, d). This ordering weakly exists in the relaxor phase and disappears in the ferroelectric phase. Because the intensity is proportional to the ordering of domains in electron diffraction [22], the weak intensities of {½ ½ ½} reflections show short-range order of Yb and Nb atoms in the $x = 0.4$ relaxor phase. Notably, relaxor behaviors appear in some Pb-based perovskites when the $B$ site of the $ABO_3$ structure has random cation distributions (disordered specimens): In $Pb(Sc_{1/2}Nb_{1/2})O_3$ and $Pb(Sc_{1/2}Ta_{1/2})O_3$, ordering due to annealing leads to the suppression of relaxor behaviors in favor of ferroelectric phases while ordering leads to an antiferroelectric phase in $Pb(In_{1/2}Nb_{1/2})O_3$ [23]. The short-range cation orders play an important role in the emergence of relaxor properties via internal random electric fields in $Pb(Mg_{1/3}Nb_{2/3})O_3$ [12,24]. The present study confirmed short-range ordering of the cations in the relaxor phase of PYN–$x$PT.

We identified the ferroelectric domains related to the relaxor property (*i.e.*, the polar nanoregions) in PYN–0.4PT. The results are displayed in Fig. 4. The butterfly-shaped diffuse scattering in the electron diffraction pattern (Fig. 4a) reveals the presence of anisotropic-shaped polarized domains. Such indicative patterns of polar nanoregions [25] have been observed in other relaxor ferroelectrics [26–28]. Thus, the local structure of PYN–0.4PT was visually derived from a systematic collection of dark-field images (Fig. 4b–e). In these images, each reflection used in the dark-field imaging was visualized for breakdown of Friedel's law under two-beam excitation [29,30]. Under this condition, the ferroelectric domains parallel to the reflection appear as bright domains [31,32]. Figure 4(b) presents the dark-field image of the $1\bar{1}0$ reflection, which contrasts the ferroelectric domains at the nanoscale. The arrowheads in the $\bar{1}10$ images point to sites having the same contrast in Fig. 4(b) and (c), demonstrating that PYN–0.4PT possesses polar nanoregions with $[1\bar{1}0]_{pc}$ polarization directions. A similar imaging was applied to the 002 and $00\bar{2}$ reflections (Fig. 4(d) and (e)). Same-contrast sites appeared in the dark-field images of the 002 and $00\bar{2}$ reflections, suggesting that $[001]_{pc}$ components of spontaneous polarization also existed in the polar nanoregions. Although these dark-field images demonstrate the spontaneous polarization of both $[110]_{pc}$ and $[001]_{pc}$ components in the local structure of the polar nanoregions, the magnitudes of these components depended on location. Thus, both the magnitudes and directions of the polarization vectors vary within the $[110]_{pc}$ and $[001]_{pc}$ directions, suggesting a monoclinic symmetry of the polar nanoregions. Besides, some V-shaped polar nanoregions were observed in the same grain (see Supplementary Fig. 5), which agrees with the presence of the butterfly-shaped diffraction spots.

The thermal evolution of the polar nanoregions was clarified through *in situ* heating experiments. The temperature-dependent dynamics of these nanoregions are possibly linked to the temperature dependence of the dielectric properties, which display broad peaks and ferroelectric relaxation. The images in the supplementary video and panels (a)–(k) of Fig. 5 were obtained from the $1\bar{1}0$ reflections and depict the changes in the polar nanoregions at temperatures between 300 and 700 K during the heating and cooling processes. At 300 K (room temperature), the polar nanoregions exhibited nanoscale contrast. Heating the specimen reduced the size of the polar nanoregions and gradually diminished the contrast. The dielectric constant peaked at $T_m$ = 575 K (Fig. 5(l)). As revealed in the video and the snapshots in Fig. (5), the size of the polar nanoregions began decreasing upon heating through $T_m$. The contrast almost disappeared at 680 K, which is close to the estimated Burns temperature $T_B$ (i.e., the temperature at which polar nanoregions start to emerge). As the specimen was cooled from 700 K, the contrast of the polar nanoregions reappeared at 680 K and both the number and size of the polar nanoregions increased with further cooling to room temperature. At room temperature, the domains of the polar nanoregions were similar to those before heating but with a different pattern, revealing that the domain patterns of polar nanoregions are irreversibly changed by annealing.

Besides, Fig. 5a–b shows long-range band-shaped domains along with polar nanoregions at room temperature. This reveals a hierarchical ferroelectric structure in the relaxor phase. The long-range macroscopic domains disappeared at 550 K although polar nanoregions existed. Considering that the long-range domains vanished at 550 K, the macroscopic domains should be caused by long-range ferroelectric interactions with the energy comparable to 550 K.

These results excellently agree with the temperature-dependent permittivity and x-ray diffraction measurements (see Fig. 6). The derivative of the inverse dielectric permittivity $\xi$ was constant at temperatures above approximately 700 K and exhibited a linear trend at lower temperatures. The onset of the relaxor properties matched the structural signature of relaxor properties in the TEM observations. Furthermore, slight distortions in these polar nanoregions were observed in the x-ray diffraction patterns. Figure 6 shows the thermal evolutions of the 002 and 111 reflections. The 002 reflection began splitting at the temperature $T_B$. The behavior of the 002 reflection is complicated because a symmetry change was observed at temperatures between $T_B$ and $T_m$. We ascribe the change to the development of dynamical polar nanoregions at $T_B$. These regions could be large or static enough to cause a structural change observable by an averaged technique of x-ray diffraction. The dynamical nature of the polar nanoregions is confirmed by the deviation from the Curie-Weiss law of the dielectric permittivity between $T_B$ and $T_m$. The 111 reflection presented an angle distortion of the lattice $\beta$ at $T_m$, suggesting a phase transition toward monoclinic phases at $T_m$. Hence, we provide direct evidence of polar nanoregions in PYN–0.4PT and their correlations with a dielectric signature of relaxor behavior and the onset of symmetry lowering at $T_B$ and $T_m$.

The right-hand-side of Fig. 1 shows the electron diffraction patterns of PYN–0.6PT, which contains the highest titanium amount among the investigated specimens. The {110} reflections were split in this composition. Dark-field imaging based on the split reflections (marked by the red and blue arrowheads in Fig. 7) yielded 90° twin domains. Such twin formation is common in tetragonal perovskites with *P4mm* symmetry, such as PbTiO$_3$ and BaTiO$_3$. The twin domains were sized 50–200 nm, demonstrating a ferroelectric composition.

To further reveal the polar character of PYN–*x*PT, the local structures were constructed via a PDF analysis, which extracts the short-range and long-range structures containing important information on polar ordering from relaxor to ferroelectric. The analysis results of the local structures are shown in Fig.

8. The long- and short-range orders varied among the compositions (Fig. 8(a) and (b)). In the antiferroelectric phase ($x = 0$), the long-range fitting result (20–100 Å) also satisfied the short-range order, suggesting that the average and local structures were the same. The PDF results support the antiferroelectric *Pmna* structure of Fig. 8(c), consistent with the HAADF–STEM results.

In the relaxor phase ($x = 0.4$), none of the long-range-fitted models matched the short-range PDF, indicating differences between the average and local structures during this phase. This result is reasonable because the local structure comprised polar nanoregions (see Fig. 4). Thus, the average and local structures were constructed by separately fitting the long and short ranges. Here, the experimental PDF was fitted by a two-phase model of *Pm* and *Cm* monoclinic phases, which could explain the average structure in this material [33] (other models such as *R3m* and *R3c* obtained inferior reliability factors in the refinement). These monoclinic structures of the relaxor phase could describe the long-range PDF in Fig. 8(a) with a reliability factor of 10.88%. The local structures were constructed by distorting the average structures. The fitting result of Fig. 8(b) produced the local structures of *Pm* and *Cm* (see Fig. 8(c)). The *Pm* and *Cm* phases were electrically polarized in the $(010)_{pc}$ and $(110)_{pc}$ planes, respectively. In the long-range fitting, the angle deviation $\beta$ is 90.2° in the *Pm* structure, demonstrating a small distortion. Hence, the average structure of the relaxor phase can be regarded as a pseudocubic structure. Conversely, the short-range fitting shows a larger distortion of $\beta = 93.7°$ in the local *Pm* structure. Thus, PYN-0.4PT has the distorted monoclinic structures in the short range and a nearly cubic structure in the long range. Notably, single-phase models of *Pm* or *Cm* structures failed to reproduce the peak shapes around 4.5 Å (Supplementary Fig. 6); furthermore, the two-phase model obtained the lowest reliability factor (Supplementary Table 1), supporting the coexistence of both monoclinic phases.

A further analysis was conducted on the ferroelectric phase of $x = 0.6$. The experimental PDF was consistent with the coexistence of the tetragonal *P4mm* and monoclinic *Pm* structures. These phases were selected because the reliability factor was lowered after assuming a monoclinic *Pm* phase. As shown in Fig. 8, the models well explained the experimental PDF in both the long and short ranges. The fractions of *P4mm* and *Pm* were approximately 62% and 38%, respectively. The main phase, *P4mm*, corresponded to the presence of tetragonal twins (see Fig. 7). However, the *P4mm* structure alone could not reproduce the experimental PDF at 4.5 Å in the short range (Supplementary Fig. 6). In the short-range, the *Pm* structure gave a lower reliability factor than *P4mm* (Supplementary Table 2). These results indicate the coexistence of *P4mm* and *Pm* structures in the ferroelectric phase of $x = 0.6$; moreover, the distortion of the local structure corresponds to *Pm* symmetry.

**Discussion**

The two-phase models allowed the fitting of the PDF results for PYN-0.4PT and PYN-0.6PT. Previous x-ray and neutron diffraction studies have confirmed phase coexistence in other perovskite solid solutions. For example, in Pb(Zr, Ti)$O_3$, rhombohedral (*R3c* or *R3m*) + monoclinic (*Cm*) structures exist at the zirconium-rich side of the phase diagram and tetragonal (*P4mm*) + monoclinic (*Cm*) structures appear in the ferroelectric phase [34–36]. A similar phase coexistence was demonstrated in ferroelectric Pb(Mg$_{1/3}$, Nb$_{2/3}$)$O_3$–PbTiO$_3$ [37,38]. A phenomenological description of higher-order Devonshire theory predicts the existence of three monoclinic phases: $M_A$ corresponding to *Cm* with a polarization vector between [111] and [001], $M_B$ corresponding to *Cm* with a polarization vector between [110] and [111], and $M_C$ corresponding to *Pm* [39]. In the theoretical phase diagram, monoclinic *Cm* and *Pm* phases are energetically degenerated and can form a mixture, consistent with the experimental results of PYN–*x*PT. Our experimental results, along with the wealth of experimental and theoretical data presented in the literature, affirm that phase coexistence with a monoclinic phase is a common characteristic near the MPB of perovskites. We note that the long-range (average) structure of PYN–*x*PT at the low PT content side of the phase diagram is not rhombohedral, contrary to all other reported Pb-based perovskite solid solutions. That is, the PYN–PT solid solution is a unique member of the Pb-based relaxor perovskite family.

The constructed structures displayed interesting trends. Figure 8(d) compares the electric polarizations in structures with different $x$ contents. The antiferroelectric displacements of the Pb atoms were parallel to $[110]_{pc}$ in the *Pmna* phase, whereas in the *P4mm* phase of $x = 0.6$, the polarization was displaced along $[001]_{pc}$. Relaxor-phase *Pm* and *Cm* could become electrically polarized in the $(010)_{pc}$ and $(110)_{pc}$ planes, the intermediate directions of the end members. Thus, the piezoelectricity enhancement originates from the monoclinic *Cm and Pm* phases, which facilitate the free rotation of polarization within the mirror planes [40–42].

Furthermore, the peculiar nature of PYN–$x$PT can be clarified by comparing its physical properties and domain structures with those of $Pb(Mg_{1/3}, Nb_{2/3})O_3$–$x$PbTiO$_3$, in which the relaxor phase ($x = 0$) has the average cubic structure ($Pm\bar{3}m$) and the local rhombohedral structure ($R3m$) while the MPB ($x = 0.35$) possesses a monoclinic *Cm* structure with high piezoelectricity [43–46]. As $x$ increases, the relaxational character of $\Delta T_{max} = T_{max}(10^6 \text{ Hz}) - T_{max}(10^2 \text{ Hz})$ decreases and reaches zero at $x = 0.35$ [12]. Conversely, the electrochemical coupling increases with $x$ and is maximized at $x = 0.35$. However, PYN–$x$PT exhibited both the relaxor property and monoclinic-associated high piezoelectricity in the relaxor phase ($0.2 < x < 0.5$; see Fig. 8e). This finding is attributable to monoclinic polar nanoregions formed by substituting Ti at the *B* sites of PYN, as demonstrated in this study. The polar nanoregions can be dynamically changed by changing the temperature through the $T_m$, as observed in the *in situ* observation: The size of polar nanoregions significantly increases with decreasing the temperature. This behavior explains the relaxor properties of dielectric dispersion because large polar nanoregions cannot follow high-frequency electric fields [47]. Moreover, monoclinic structures facilitate the rotation of electric polarization, similarly to $Pb(Mg_{1/3}, Nb_{2/3})O_3$–PbTiO$_3$ and non-relaxor $Pb(Zr, Ti)O_3$ near the MPB. The simultaneous realization of monoclinic structures and polar nanoregions related to dielectric dispersion should underlie the relaxor characteristics and high piezoelectricity in the relaxor phase. Furthermore, the domain structures of PYN–$x$PT differ from those of $Pb(Mg_{1/3}, Nb_{2/3})O_3$–PbTiO$_3$.

Whereas Pb(Mg$_{1/3}$, Nb$_{2/3}$)O$_3$–PbTiO$_3$ presents lamella-like ferroelectric nanodomains that elongate along one direction in the monoclinic phase near the MPB [48,49], PYN–xPT presented a random distribution of polar nanoregions (Fig. 4). This difference might also contribute to dielectric dispersion because the relaxor property is attributable to random orientations of electric polarization, *i.e.*, inner random electric fields, caused by inhomogeneous cation distributions.

**Conclusions**

This study revealed unprecedented polar structures of monoclinic polar nanoregions. A comprehensive local structural analysis of PYN–xPT also revealed unique structural changes in this material. The antiferroelectric phase ($x = 0$ and 0.05) exhibited transverse ionic modulations of (3/8, 3/8, 0)$_{pc}$ along the [110]$_{pc}$ direction. The displacements of these modulations were visualized using atomic-resolution STEM. Cation ordering induced antiphase boundaries in the $x = 0.2$ compound. As the PbTiO$_3$ content increased, the electric-polarization direction changed from [110]$_{pc}$ to [001]$_{pc}$ through the (110)$_{pc}$ and (010)$_{pc}$ planes. Temperature-changeable monoclinic polar nanoregions were observed in the relaxor region of PYN–xPT, simultaneously enabling the relaxor property and high piezoelectricity and large electromechanical coupling factors near the MPB. This simultaneous realization is unique compared with Pb(Mg$_{1/3}$, Nb$_{2/3}$)O$_3$–PbTiO$_3$ and other relaxor-based solid solutions.

**Acknowledgments**

This study was supported in part by JSPS KAKENHI Grant Numbers JP19H05814, JP19H05625, JP20K15031, JP21K14538, and JP21H04625. The synchrotron radiation experiments were performed at BL04B2 of SPring-8 with the approval of the Japan Synchrotron Radiation Research Institute (Proposal No. 2022A1085).

**REFERENCES**

[1] J. Kuwata, K. Uchino, and S. Nomura, *Phase Transitions in the Pb (Zn$_{1/3}$Nb$_{2/3}$)O$_3$-PbTiO$_3$ System*, Ferroelectrics **37**, 579 (1981).

[2] S.-E. Park and T. R. Shrout, *Ultrahigh Strain and Piezoelectric Behavior in Relaxor Based Ferroelectric Single Crystals*, J. Appl. Phys. **82**, 1804 (1997).

[3] G. Burns and F. H. Dacol, *Glassy Polarization Behavior in Ferroelectric Compounds Pb(Mg$_{1/3}$Nb$_{2/3}$)O$_3$ and Pb(Zn$_{1/3}$Nb$_{2/3}$)O$_3$*, Solid State Commun. **48**, 853 (1983).

[4] A. A. Bokov and Z.-G. Ye, *Recent Progress in Relaxor Ferroelectrics with Perovskite Structure*, J. Mater. Sci. **41**, 31 (2006).

[5] F. Li et al., *The Origin of Ultrahigh Piezoelectricity in Relaxor-Ferroelectric Solid Solution Crystals*, Nat. Commun. **7**, 1 (2016).

[6] T. Yamamoto and S. Ohashi, *Dielectric and Piezoelectric Properties of Pb(Yb$_{1/2}$Nb$_{1/2}$)O$_3$-PbTiO$_3$ Solid Solution System*, Jpn. J. Appl. Phys. **34**, 5349 (1995).

[7] S. Zhang, S. Rhee, C. A. Randall, and T. R. Shrout, *Dielectric and Piezoelectric Properties of High Curie Temperature Single Crystals in the Pb(Yb$_{1/2}$Nb$_{1/2}$)O$_3$-xPbTiO$_3$ Solid Solution Series*, Jpn. J. Appl. Phys. **41**, 722 (2002).

[8] S. Zhang, P. W. Rehrig, C. Randall, and T. R. Shrout, *Crystal Growth and Electrical Properties of Pb(Yb$_{1/2}$Nb$_{1/2}$)O$_3$-PbTiO$_3$ Perovskite Single Crystals*, J. Cryst. Growth **234**, 415 (2002).

[9] J. B. Lim, S. Zhang, and T. R. Shrout, *Relaxor Behavior of Piezoelectric Pb(Yb$_{1/2}$Nb$_{1/2}$)O$_3$-PbTiO$_3$ Ceramics Sintered at Low Temperature*, J. Electroceramics **26**, 68 (2011).

[10] H. Lim, H. J. Kim, and W. K. Choo, *X-Ray and Dielectric Studies of the Phase Transitions in Pb(Yb$_{1/2}$Nb$_{1/2}$)O$_3$-PbTiO$_3$ Ceramics*, Jpn. J. Appl. Phys. **34**, 5449 (1995).

[11] C. Cochard, X. Bril, O. Guedes, and P.-E. Janolin, *Interpretation of Polar Orders Based on Electric Characterizations: Example of Pb(Yb$_{1/2}$Nb$_{1/2}$)O$_3$-PbTiO$_3$ Solid Solution*, J. Electron. Mater. **45**, 6005 (2016).

[12] M. J. Krogstad et al., *The Relation of Local Order to Material Properties in Relaxor Ferroelectrics*, Nat. Mater. **17**, 718 (2018).

[13] T. E. Faber and J. M. Ziman, *A Theory of the Electrical Properties of Liquid Metals: III. The Resistivity of Binary Alloys*, Philos. Mag. **11**, 153 (1965).

[14] S. Hiroi, K. Ohara, S. Ohuchi, Y. Umetani, T. Kozaki, E. Igaki, and O. Sakata, *Calculation of Total Scattering from a Crystalline Structural Model Based on Experimental Optics Parameters*, J. Appl. Crystallogr. **53**, 671 (2020).

[15] W. K. Choo, H. J. Kim, J. H. Yang, H. Lim, J. Y. Lee, J. R. Kwon, and C. H. Chun, *Crystal Structure and B-Site Ordering in Antiferroelectric Pb(Mg$_{1/2}$W$_{1/2}$)O$_3$, Pb(Co$_{1/2}$W$_{1/2}$)O$_3$ and Pb(Yb$_{1/2}$Nb$_{1/2}$)O$_3$*, Jpn. J. Appl. Phys. **32**, 4249 (1993).

[16] J. R. Kwon and W. K. Choo, *The Antiferroelectric Crystal Structure of the Highly Ordered Complex Perovskite Pb(Yb$_{1/2}$Nb$_{1/2}$)O$_3$*, J. Phys. Condens. Matter **3**, 2147 (1991).

[17] J. R. Kwon, C. K. K. Choo, and W. K. Choo, *Dielectric and X-Ray Diffraction Studies in Highly Ordered Complex Perovskite Pb(Yb$_{1/2}$Nb$_{1/2}$)O$_3$*, Jpn. J. Appl. Phys. **30**, 1028 (1991).

[18] Z. Chen, X. Wang, S. P. Ringer, and X. Liao, *Manipulation of Nanoscale Domain Switching Using an Electron Beam with Omnidirectional Electric Field Distribution*, Phys. Rev. Lett. **117**, 27601 (2016).

[19] Y. Wang, F.-T. Huang, X. Luo, B. Gao, and S.-W. Cheong, *The First Room-Temperature Ferroelectric Sn Insulator and Its Polarization Switching Kinetics*, Adv. Mater. **29**, 1601288 (2017).

[20] C. A. Randall and A. S. Bhalla, *Nanostructural-Property Relations in Complex Lead Perovskites*, Jpn. J. Appl. Phys. **29**, 327 (1990).

[21] A. D. Hilton, D. J. Barber, C. A. Randall, and T. R. Shrout, *On Short Range Ordering in the Perovskite Lead Magnesium Niobate*, J. Mater. Sci. **25**, 3461 (1990).

[22] J. Chen, H. M. Chan, and M. P. Harmer, *Ordering Structure and Dielectric Properties of Undoped and La/Na-Doped Pb($Mg_{1/3}Nb_{2/3}$)$O_3$*, J. Am. Ceram. Soc. **72**, 593 (1989).

[23] N. Setter and L. E. Cross, *The Contribution of Structural Disorder to Diffuse Phase Transitions in Ferroelectrics*, J. Mater. Sci. **15**, 2478 (1980).

[24] S. Prosandeev and L. Bellaiche, *Effects of Atomic Short-Range Order on Properties of the $PbMg_{1/3}Nb_{2/3}O_3$ Relaxor Ferroelectric*, Phys. Rev. B **94**, 180102 (2016).

[25] M. Eremenko, V. Krayzman, A. Bosak, H. Y. Playford, K. W. Chapman, J. C. Woicik, B. Ravel, and I. Levin, *Local Atomic Order and Hierarchical Polar Nanoregions in a Classical Relaxor Ferroelectric*, Nat. Commun. **10**, 1 (2019).

[26] M. Matsuura, K. Hirota, P. M. Gehring, Z.-G. Ye, W. Chen, and G. Shirane, *Composition Dependence of the Diffuse Scattering in the Relaxor Ferroelectric Compound $(1-x)Pb(Mg_{1/3}Nb_{2/3})O_3 - xPbTiO_3$ ($0 \leq X \leq 0.40$)*, Phys. Rev. B **74**, 144107 (2006).

[27] G. Xu, Z. Zhong, H. Hiraka, and G. Shirane, *Three-Dimensional Mapping of Diffuse Scattering in $Pb(Zn_{1/3}Nb_{2/3})O_3-XPbTiO_3$*, Phys. Rev. B **70**, 174109 (2004).

[28] P. M. Gehring, H. Hiraka, C. Stock, S.-H. Lee, W. Chen, Z.-G. Ye, S. B. Vakhrushev, and Z. Chowdhuri, *Reassessment of the Burns Temperature and Its Relationship to the Diffuse Scattering, Lattice Dynamics, and Thermal Expansion in Relaxor $Pb(Mg_{1/3}Nb_{2/3})O_3$*, Phys. Rev. B **79**, 224109 (2009).

[29] F. Fujimoto, *Dynamical Theory of Electron Diffraction in Laue-Case, I. General Theory*, J. Phys. Soc. Japan **14**, 1558 (1959).

[30] M. Tanaka and G. Honjo, *Electron Optical Studies of Barium Titanate Single Crystal Films*, J. Phys. Soc. Japan **19**, 954 (1964).

[31] T. Asada and Y. Koyama, *Ferroelectric Domain Structures around the Morphotropic Phase Boundary of the Piezoelectric Material $PbZr_{1-x}Ti_xO_3$*, Phys. Rev. B **75**, 214111 (2007).

[32] H. Tsukasaki, Y. Uneno, S. Mori, and Y. Koyama, *Features of Ferroelectric States in the Simple-Perovskite Mixed-Oxide System $(1-x)Pb(Zn_{1/3}Nb_{2/3})O_3-xPbTiO_3$ with Lower Ti Contents*, J. Phys. Soc. Japan **85**, 34708 (2016).

[33] C. Cochard, $Pb(Yb_{1/2}Nb_{1/2})_3$-$PbTiO_3$: A Model Solid Solution for the Study of the Different Polar Orders, Châtenay-Malabry, Ecole centrale de Paris, 2015.

[34] N. Zhang, H. Yokota, A. M. Glazer, Z. Ren, D. A. Keen, D. S. Keeble, P. A. Thomas, and Z.-G. Ye, *The Missing Boundary in the Phase Diagram of $PbZr_{1-x}Ti_xO_3$*, Nat. Commun. **5**, 1 (2014).

[35] H. Yokota, N. Zhang, A. E. Taylor, P. A. Thomas, and A. M. Glazer, *Crystal Structure of the Rhombohedral Phase of $PbZr_{1-x}Ti_xO_3$ Ceramics at Room Temperature*, Phys. Rev. B **80**, 104109 (2009).

[36] N. Zhang, H. Yokota, A. M. Glazer, and P. A. Thomas, *Neutron Powder Diffraction Refinement of $PbZr_{1-x}Ti_xO_3$*, Acta Crystallogr. Sect. B Struct. Sci. **67**, 386 (2011).

[37] Z.-G. Ye, B. Noheda, M. Dong, D. Cox, and G. Shirane, *Monoclinic Phase in the Relaxor-Based Piezoelectric/Ferroelectric $Pb(Mg_{1/3}Nb_{2/3})O_3$-$PbTiO_3$ System*, Phys. Rev. B **64**, 184114 (2001).

[38] B. Noheda, D. E. Cox, G. Shirane, J. Gao, and Z.-G. Ye, *Phase Diagram of the Ferroelectric Relaxor $(1-x)PbMg_{1/3}Nb_{2/3}O_3-xPbTiO_3$*, Phys. Rev. B **66**, 54104 (2002).

[39] D. Vanderbilt and M. H. Cohen, *Monoclinic and Triclinic Phases in Higher-Order Devonshire Theory*, Phys. Rev. B **63**, 94108 (2001).

[40]  B. Noheda, D. E. Cox, G. Shirane, J. A. Gonzalo, L. E. Cross, and S. E. Park, *A Monoclinic Ferroelectric Phase in the Pb(Zr$_{1-x}$Ti$_x$)O$_3$ Solid Solution*, Appl. Phys. Lett. **74**, 2059 (1999).

[41]  H. Fu and R. E. Cohen, *Polarization Rotation Mechanism for Ultrahigh Electromechanical Response in Single-Crystal Piezoelectrics*, Nature **403**, 281 (2000).

[42]  L. Bellaiche, A. García, and D. Vanderbilt, *Finite-Temperature Properties of Pb(Zr$_{1-x}$Ti$_x$)O$_3$ Alloys from First Principles*, Phys. Rev. Lett. **84**, 5427 (2000).

[43]  N. De Mathan, E. Husson, G. Calvarn, J. R. Gavarri, A. W. Hewat, and A. Morell, *A Structural Model for the Relaxor PbMg$_{1/3}$Nb$_{2/3}$O$_3$ at 5 K*, J. Phys. Condens. Matter **3**, 8159 (1991).

[44]  P. Bonneau, P. Garnier, G. Calvarin, E. Husson, J. R. Gavarri, A. W. Hewat, and A. Morell, *X-Ray and Neutron Diffraction Studies of the Diffuse Phase Transition in PbMg$_{1/3}$Nb$_{2/3}$O$_3$ Ceramics*, J. Solid State Chem. **91**, 350 (1991).

[45]  I.-K. Jeong, T. W. Darling, J. K. Lee, T. Proffen, R. H. Heffner, J. S. Park, K. S. Hong, W. Dmowski, and T. Egami, *Direct Observation of the Formation of Polar Nanoregions in Pb(Mg$_{1/3}$Nb$_{2/3}$)O$_3$ Using Neutron Pair Distribution Function Analysis*, Phys. Rev. Lett. **94**, 147602 (2005).

[46]  Y. Yoneda, H. Taniguchi, and Y. Noguchi, *Nanoscale Structural Analysis of Pb(Mg$_{1/3}$Nb$_{2/3}$)O$_3$*, J. Phys. Condens. Matter **33**, 35401 (2020).

[47]  Z. G. Lu and G. Calvarin, *Frequency Dependence of the Complex Dielectric Permittivity of Ferroelectric Relaxors*, Phys. Rev. B **51**, 2694 (1995).

[48]  M. Otoničar et al., *Connecting the Multiscale Structure with Macroscopic Response of Relaxor Ferroelectrics*, Adv. Funct. Mater. **30**, 2006823 (2020).

[49]  Y. Sato, S. Fujinaka, S. Yamaguchi, R. Teranishi, K. Kaneko, T. Shimizu, H. Taniguchi, and H. Moriwake, *Lamellar-like Nanostructure in a Relaxor Ferroelectrics Pb(Mg$_{1/3}$Nb$_{2/3}$)O$_3$*, J. Mater. Sci. **56**, 1231 (2021).

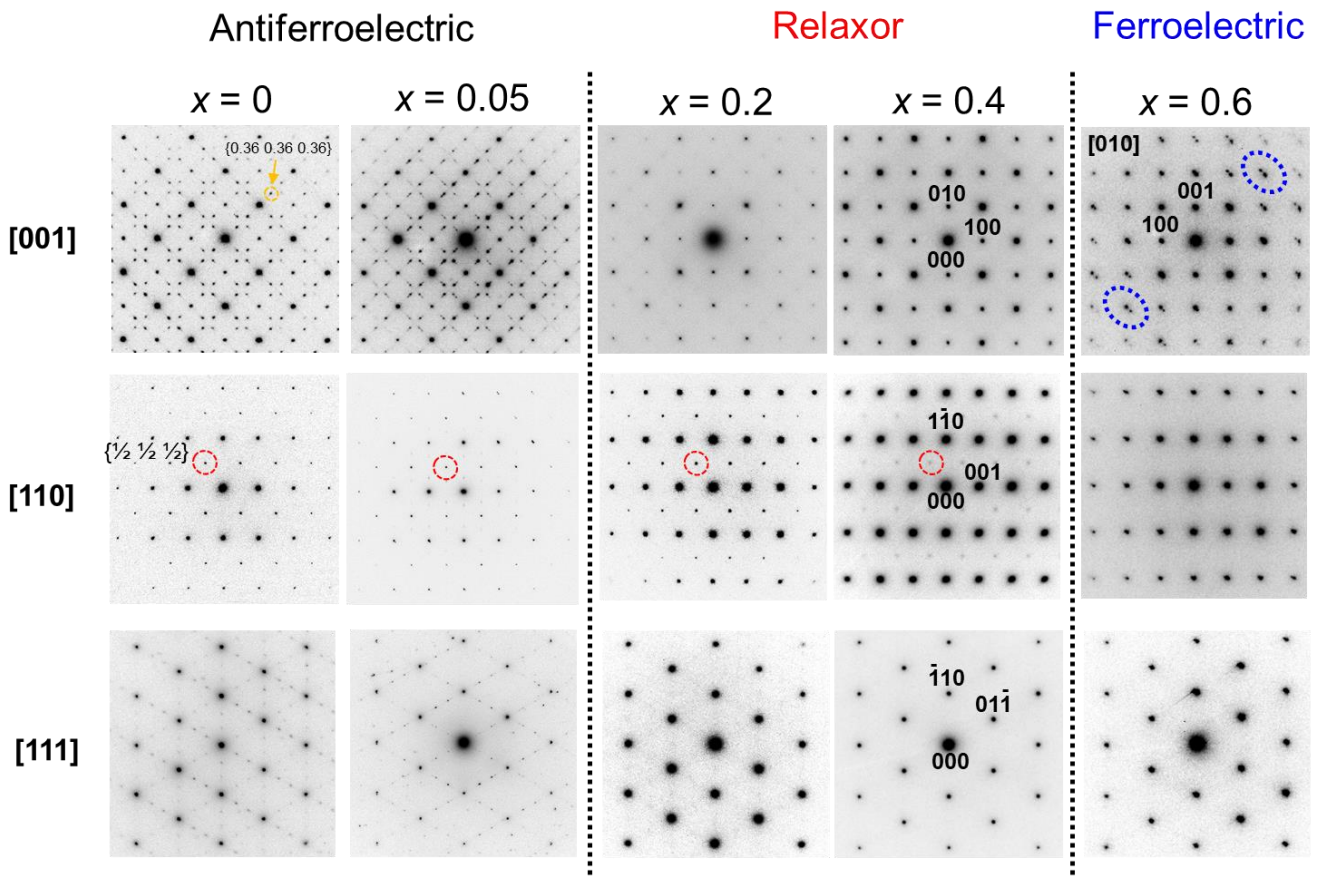

**Figure 1** Compositional dependence of the electron diffraction patterns in PbYb$_{1/2}$Nb$_{1/2}$O$_3$–$x$PbTiO$_3$. The indices are based on the pseudocubic perovskite structure of PYN–0.4PT. The red-edged circles and blue-edged ellipses exemplify {½ ½ ½}-type reflections due to cation ordering in $x$ = 0–0.4 and split reflections originating from twins in the $x$ = 0.6 phase, respectively. The selected-area diffraction patterns were obtained from areas with a diameter of 500 nm.

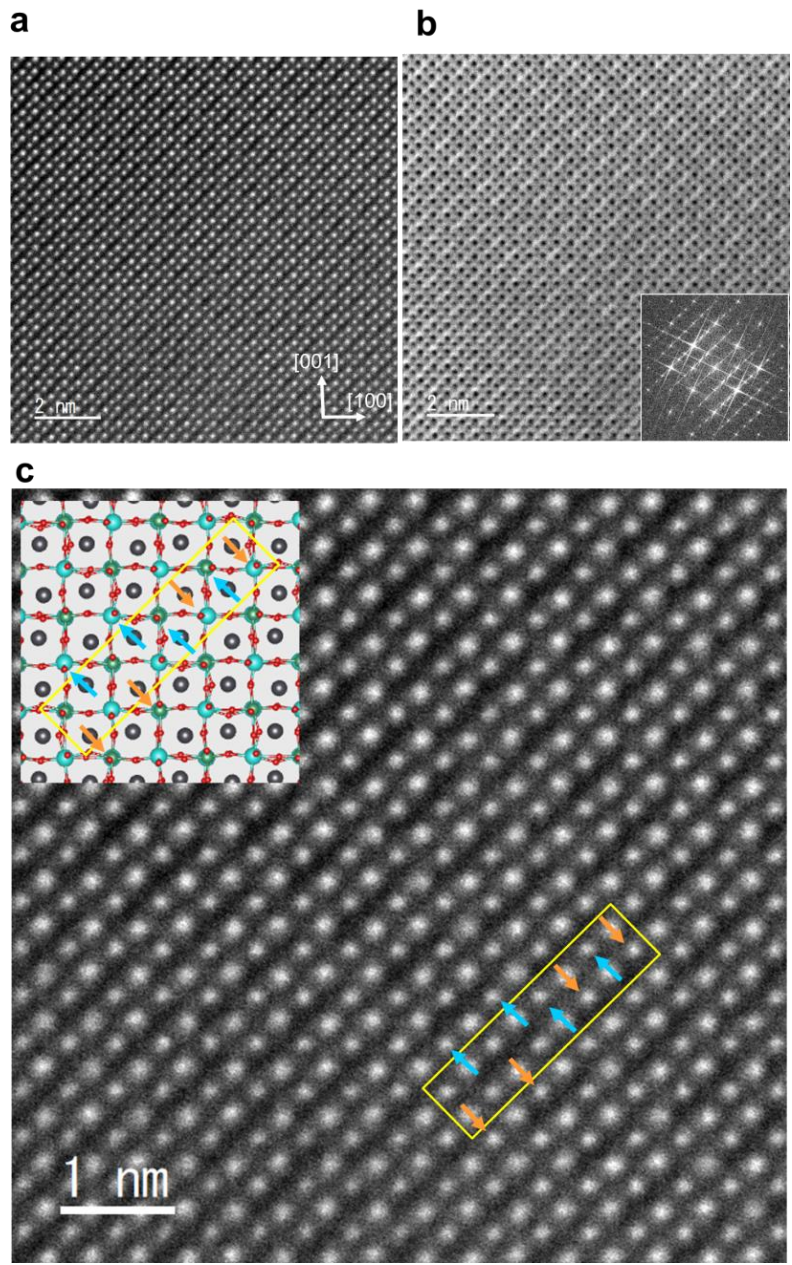

**Figure 2** Atomic-resolution STEM analysis. (a) HAADF–STEM and (b) bright-field STEM images along the [001] direction in PbYb$_{1/2}$Nb$_{1/2}$O$_3$ (PYN–0PT). The inset in (b) is the fast Fourier transform pattern. (c) High-magnification image showing antiferroelectric displacements (pointed by arrows). The yellow rectangle delineates the unit cell of the *Pmna* structure. The inset is a schematic of the crystal structure viewed along the *b* axis of *Pmna*. The gray, blue, green, and red spheres denote Pb, Yb, Nb, and O, respectively.

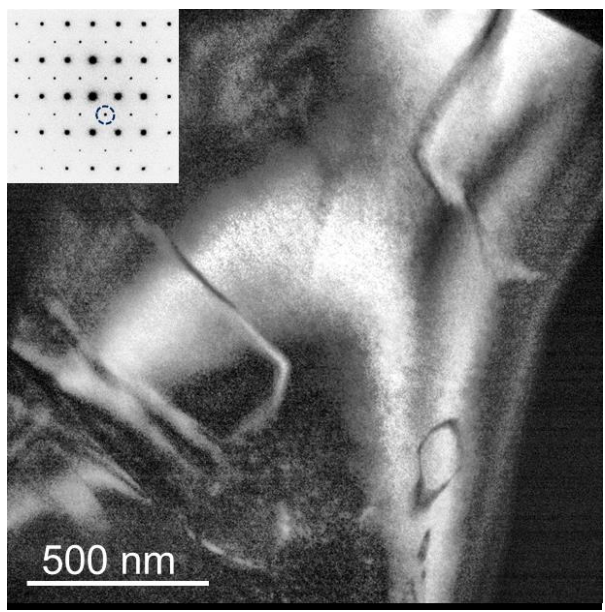

**Figure 3** Dark-field image of PYN–0.2PT showing antiphase boundaries. The inset shows the {½ ½ ½} superlattice reflection used for the dark-field image.

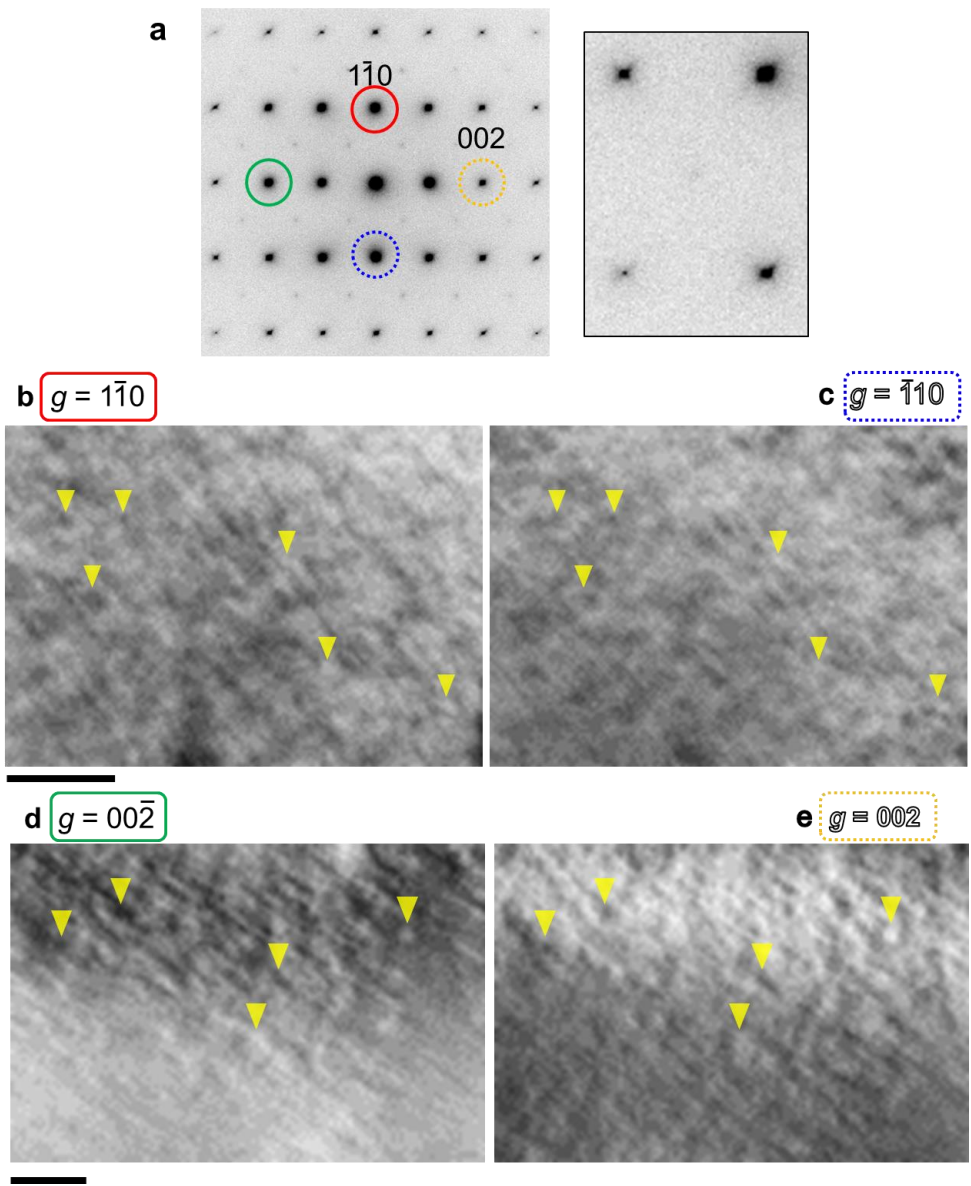

**Figure 4** Polar nanoregions in PYN–0.4PT. (a) Electron diffraction pattern along [110]. The right panel is a magnified image. (b, c) [1$\bar{1}$0] components of the polar nanoregions: Dark-field images based on the (b) 1$\bar{1}$0 and (c) $\bar{1}$10 reflections. (d, e) [001] components of the polar nanoregions: Dark-field images based on the (d) 00$\bar{2}$ and (e) 002 reflections. The contrast in images (c) and (e) was reversed by reversing the experimentally obtained intensities. In (b) and (c) and similarly in (d) and (e), the arrowheads point to the sites of equal contrast, which demonstrates the contrast reversal between the dark-field images of the $g$ and $\bar{g}$ reflections. All images were obtained on the same grain. The scale bars are 100 nm.

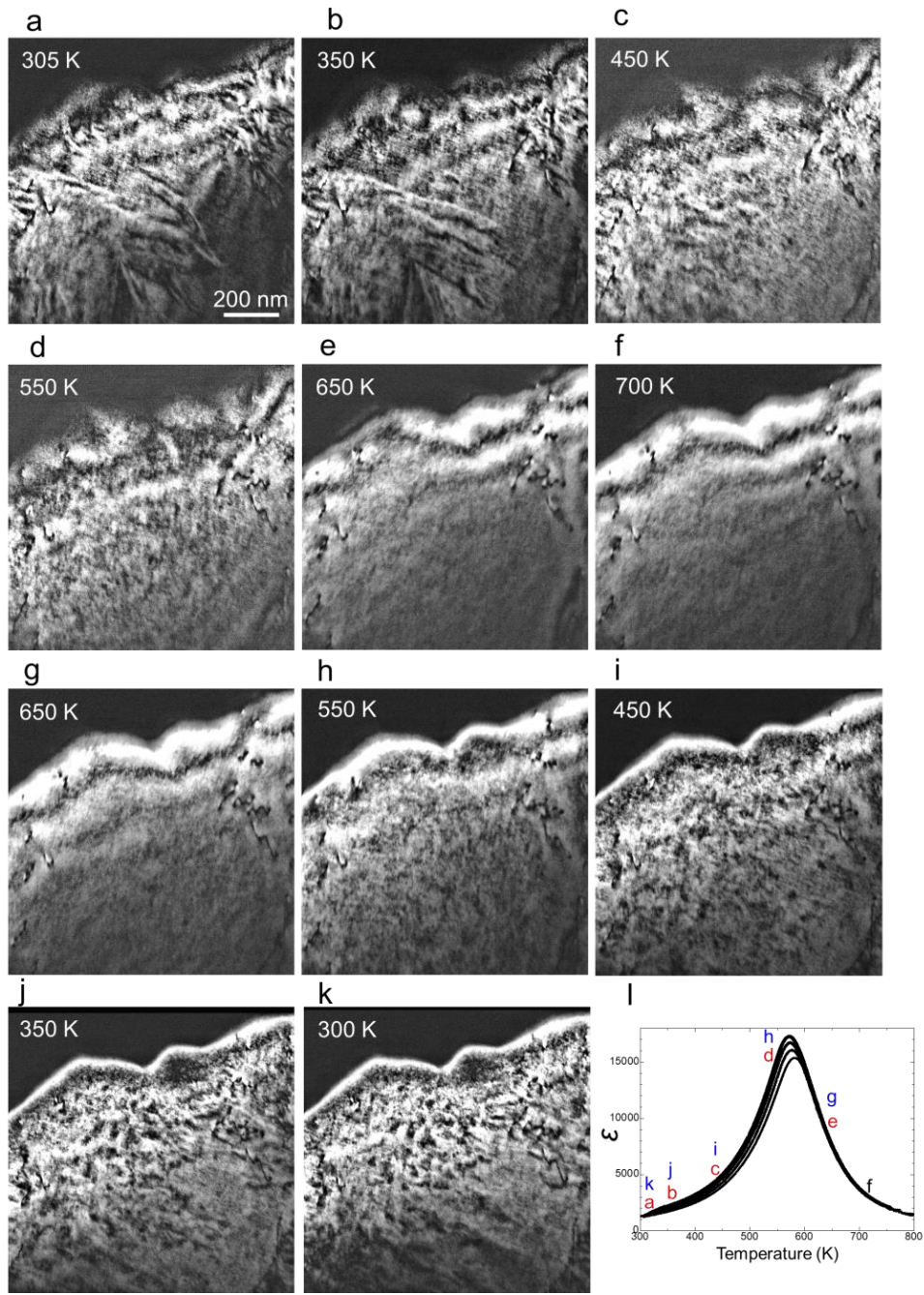

**Figure 5** *In situ* heating observation of polar nanoregions in PYN–0.4PT. (a)–(k) Dark-field images based on $1\bar{1}0$ reflection (extracted from the supplementary movie). (l) Temperature dependence of the dielectric constants in PYN–0.4PT. The alphabetical symbols correspond to the images obtained during heating (red) and cooling (blue). The permittivity data are reproduced from Ref. [11].

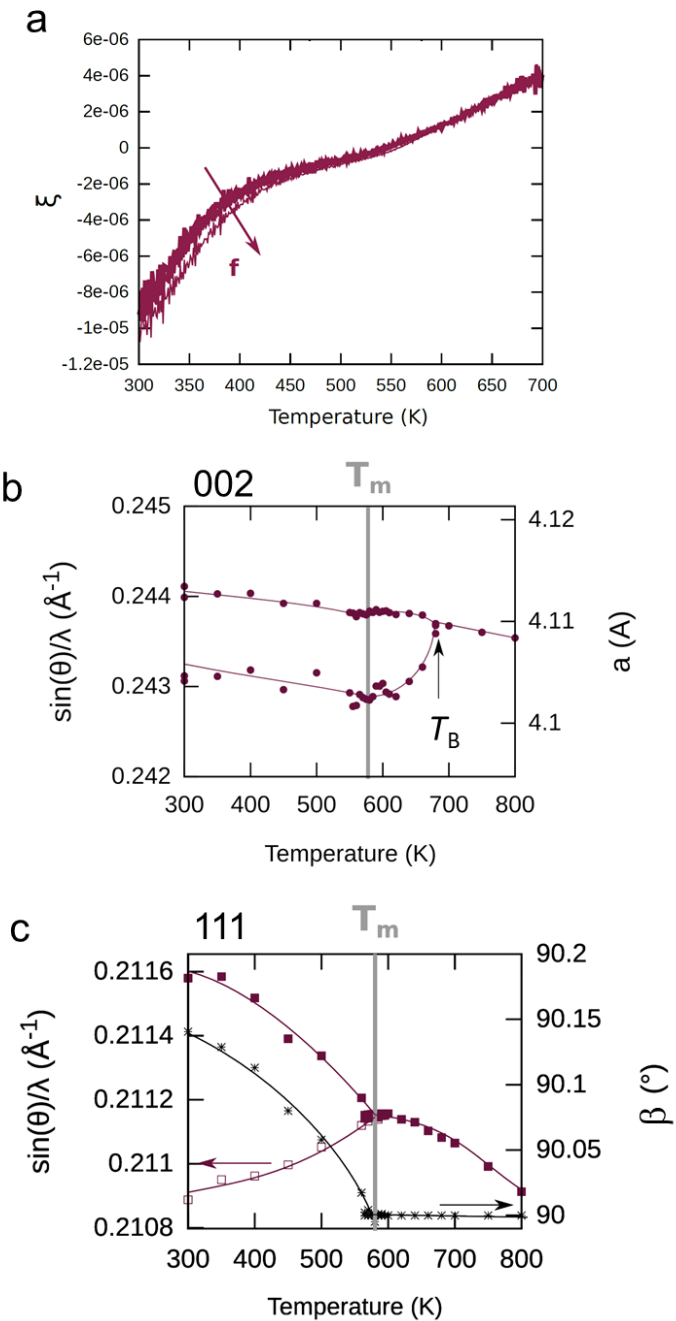

**Figure 6** Temperature evolution of permittivity and x-ray diffraction intensity in PYN–0.4PT. (a) Temperature dependences of (a) the derivative of the inverse dielectric constant $\xi$, (b) the 002 reflection, and (c) the 111 reflection. The right axes in panels (b) and (c) refer to the lattice constant $a$ and angle $\beta$ of the unit cell, respectively. $T_m$ and $T_B$ represent the temperature of the maximum dielectric constant and the Burns temperature, respectively.

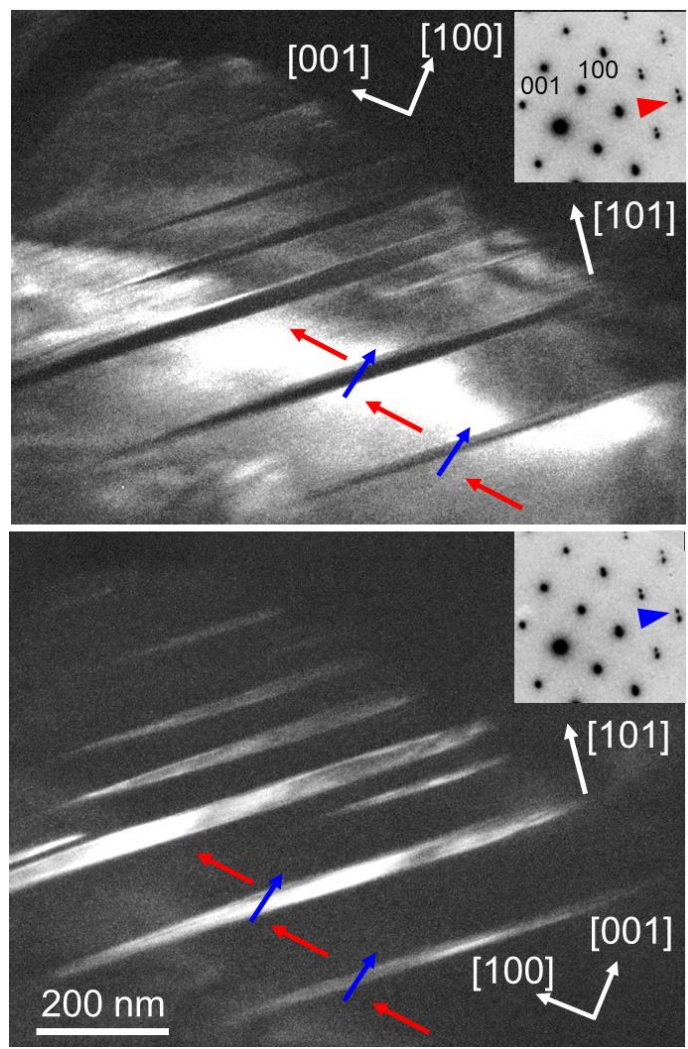

**Figure 7** Dark-field images demonstrating twin domains in PYN–0.6PT. The red and blue arrows indicate the electric-polarization directions. The insets are electron diffraction patterns indicating the used reflection.

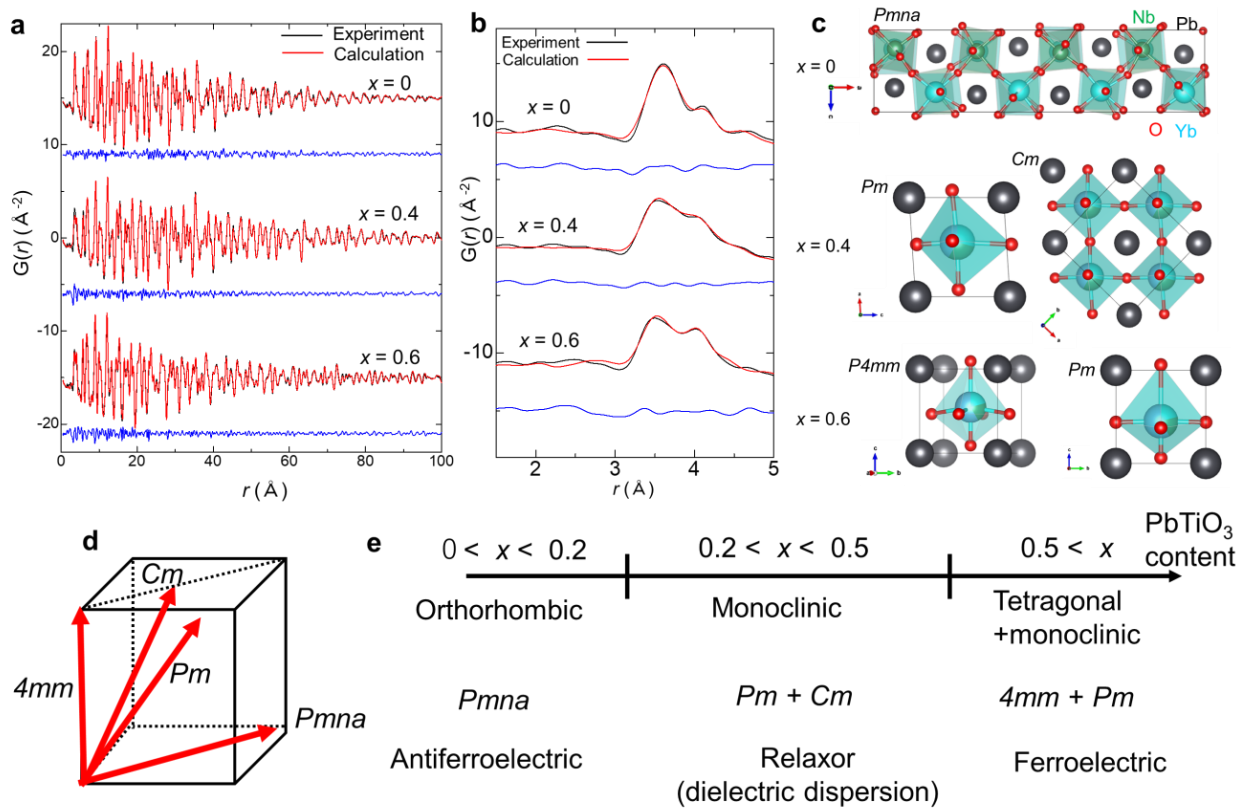

**Figure 8** Pair-distribution function (PDF) analysis of PbYb$_{1/2}$Nb$_{1/2}$O$_3$–$x$PbTiO$_3$ with $x$ = 0, 0.4, and 0.6. (a) Experimental PDF and the long-range (20–100 Å) fitting results refined by structural models. The blue lines show the differences between the experimental and calculated PDFs. (b) Fitting results in the short-range (1.5–5 Å). (c) Crystal structures after refinement. Two phases coexist in the compositions with $x$ = 0.4 and 0.6. The fitting results are listed in the Supplementary Information. (d) Electric-polarization directions in each phase. (e) Phase diagram derived in this study.

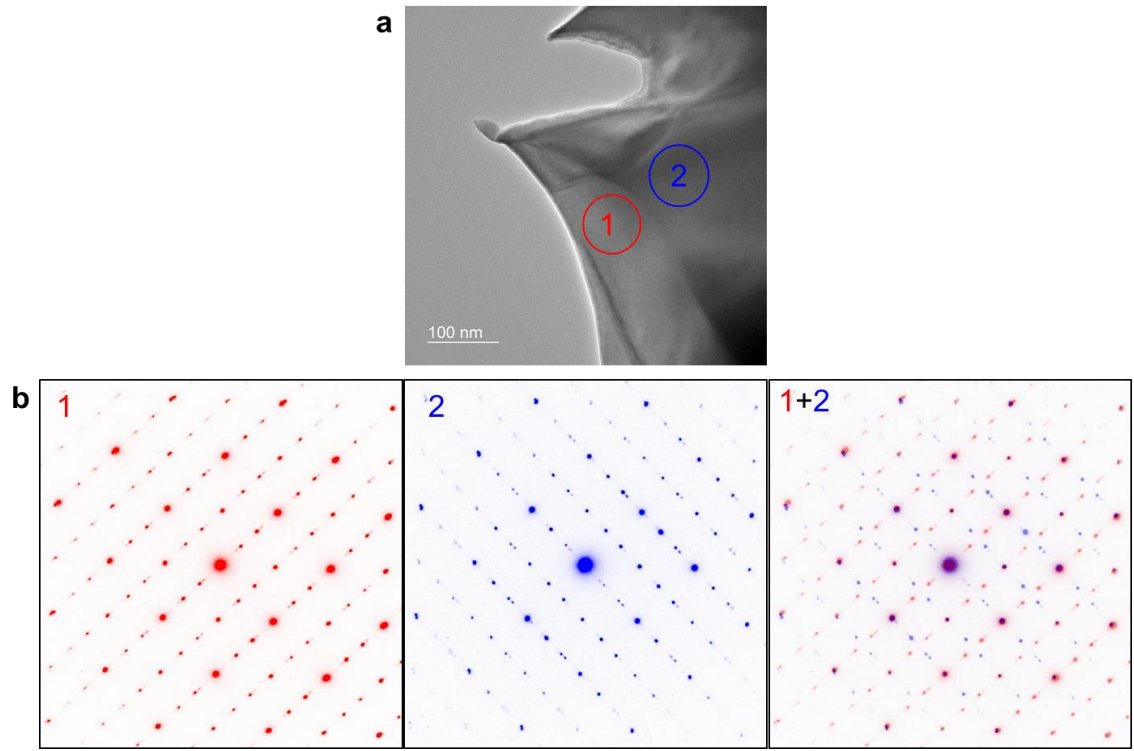

Supplementary Figure 1. Grain boundary viewed along [001] in $PbYb_{1/2}Nb_{1/2}O_3$. (a) Bright-field image. The marks are the areas for selected-area electron diffraction patterns. (b) Electron diffraction patterns in areas 1 and 2. The right panel is the superimposed patterns of diffraction patterns 1 and 2.

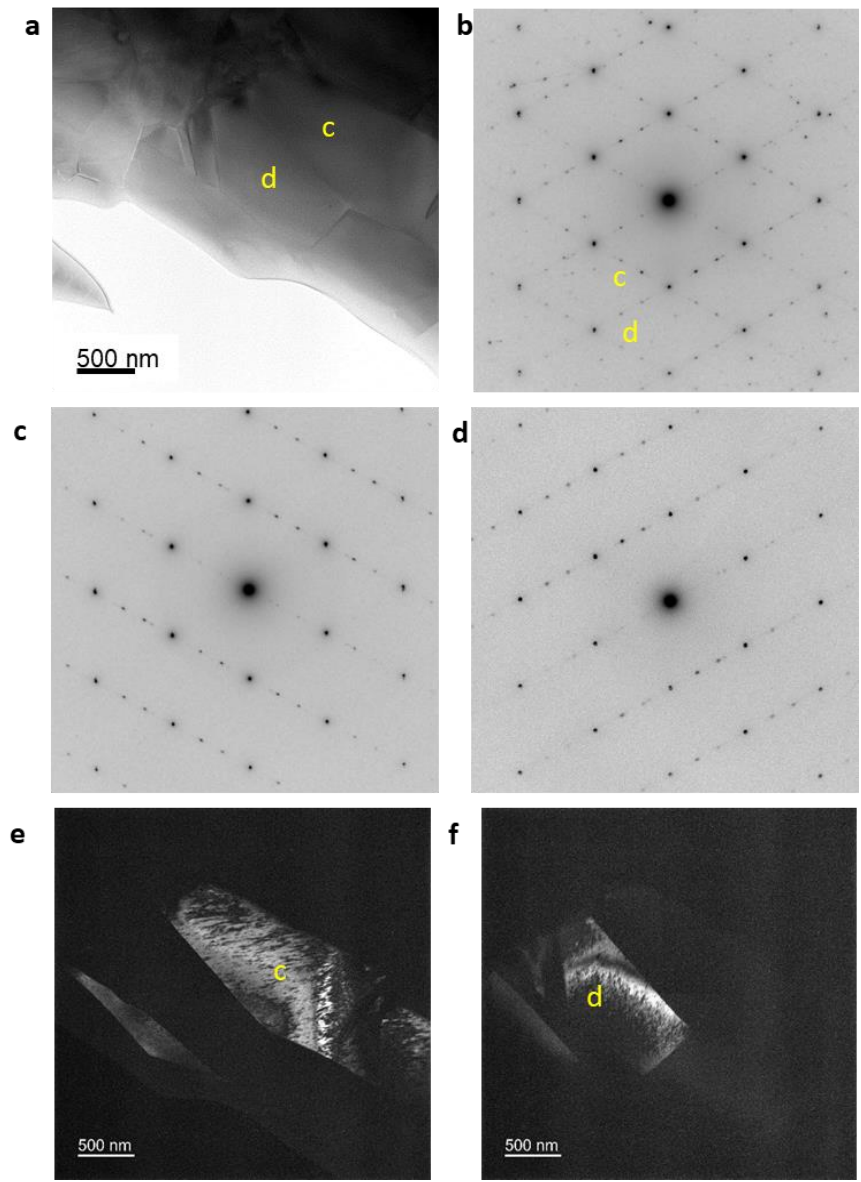

Supplementary Figure 2. Grain boundary viewed along [111] in PYN-0.05PT. (a) Bright-field image. (b) Electron diffraction pattern from both the grains *c* and *d* of panel (a). (c, d) Electron diffraction patterns from the grains (c) and (d). (e, f) Dark-field images obtained by the spot *c* and *d* of panel (b), respectively.

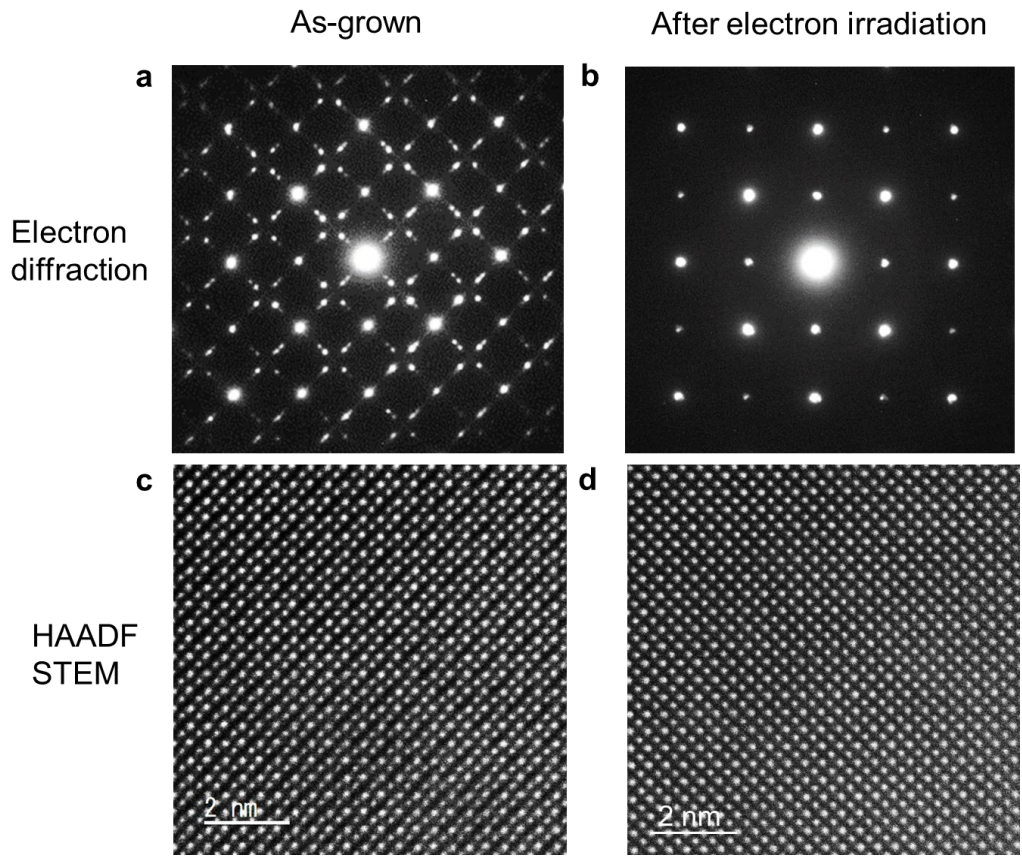

Supplementary Figure 3. Disappearance of antiferroelectric displacements. (a, b) Electron diffraction patterns (a) before and (b) after electron irradiation of 10 minutes. (c, d) HAADF-STEM images (c) before and (d) after electron irradiation. Electron irradiation changed the structure into a simple perovskite without modulation. The transition temperature $T_N$ from the paraelectric to antiferroelectric phases is 553 K. Thus, heating due to electron irradiation is negligible in the transition. This suggests that local electric fields due to electron charging caused this phenomenon.

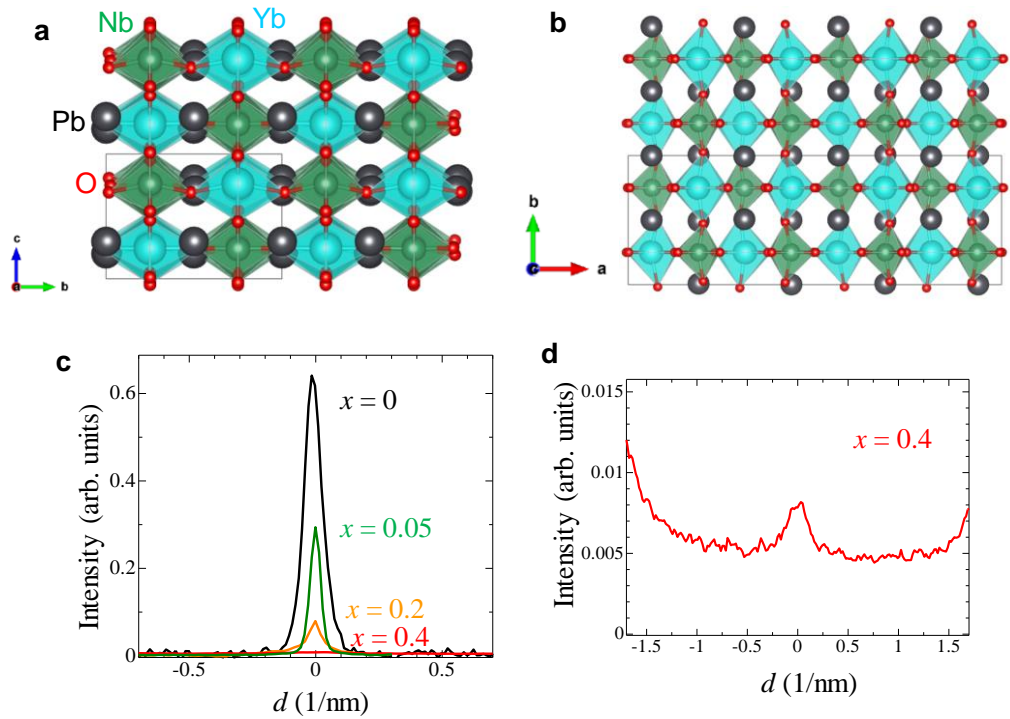

Supplementary Figure 4. {½ ½ ½} cation-ordering structure and composition dependence of the intensities of {½ ½ ½} spots. (a, b) Schematics of the *Pmna* structure along *a* and *c* axes. (c) Intensity profile of the {½ ½ ½} spot in each composition. The intensities of the {½ ½ ½} spots are normalized by those of the 111 reflections in electron diffraction patterns. (d) Magnified profile of *x* = 0.4 in panel (c).

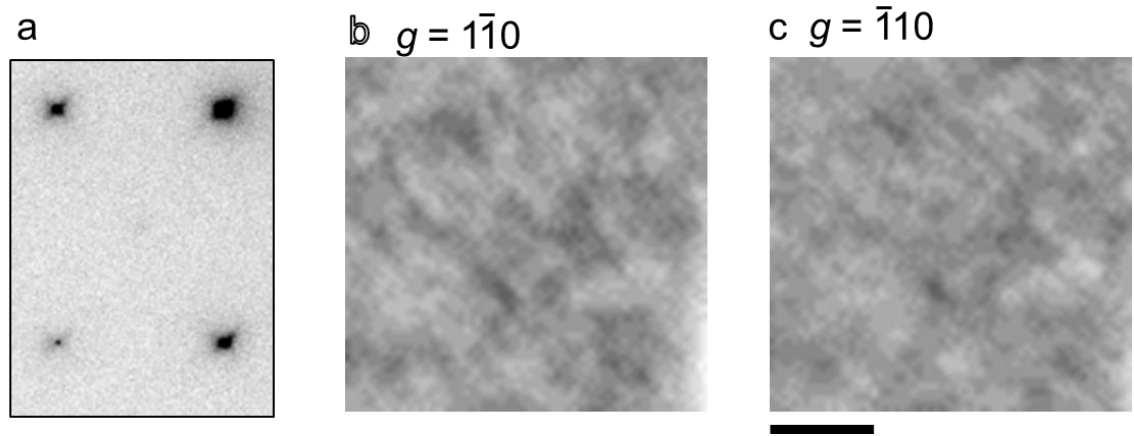

Supplementary Figure 5. The relation between butterfly-shaped spots and polar nanoregions. (a) Electron diffraction pattern. (b) Dark-field image via $1\bar{1}0$ reflection with contrast reversed. (c) Dark-field image via $\bar{1}10$ reflection. The same patterns demonstrate the contrast due to ferroelectric domains. The V-shaped domains are consistent with the presence of the butterfly-shaped spots. The scale bar is 50 nm.

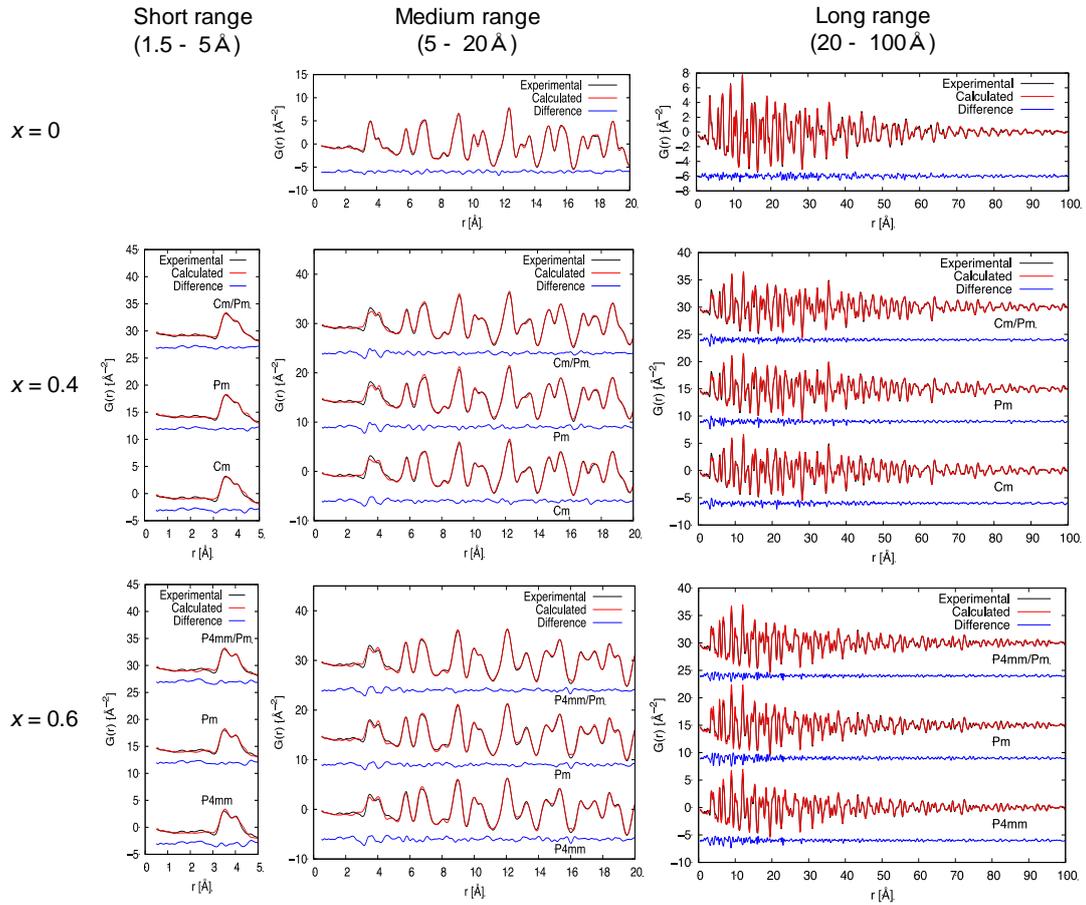

Supplementary Figure 6. Experimental PDF and fitting results in each range. In $x = 0$, the reliability factor $R_{wp} = 7.503\%$ for the medium range and $R_{wp} = 15.17\%$ for the long range. The marks $Cm/Pm$ and $P4/mm/Pm$ mean the models assuming the two phases.

Supplementary Table 1. The reliability factor $R_{wp}$ (%) of each model in PYN-0.4PT. The volume fraction of $Pm/Cm$ is approximately 53%:47%.

|  | *Pm* | *Cm* | *Pm/Cm* |
| --- | --- | --- | --- |
| Long range | 12.81% | 13.03% | 10.88% |
| Medium range | 9.236% | 8.602% | 7.475% |
| Short range | 12.12% | 16.18% | 10.87% |

Supplementary Table 2. The reliability factor $R_{wp}$ (%) of each model in PYN-0.6PT. The volume fraction of *P4mm*/*Pm* is approximately 62%:38%.

|  | *P4mm* | *Pm* | *P4mm/Pm* |
|---|---|---|---|
| Long range | 14.03% | 16.08% | 13.66% |
| Medium range | 10.42% | 8.718% | 7.963% |
| Short range | 21.92% | 16.92% | 17.75% |

Supplementary Table 3. Crystal structure of the monoclinic local structure *Pm* (two-phase model) in PYN-0.4PT. The parameters are refined to fit the experimental PDF in the long range.

|  | *Pm* |
|---|---|
| Lattice Parameters |  |
| $a$ (Å) | 4.09373 |
| $b$ (Å) | 4.08906 |
| $c$ (Å) | 4.09852 |
| $\beta$ (°) | 90.204 |
| Atomic positions ($x, y, z$) |  |
| Pb | 0, 0, 0 |
| O1 | 0.565, 0.000, 0.428 |
| O2 | 0.528, 0.500, 0.050 |
| O3 | 0.073, 0.500, 0.460 |
| Yb/Nb/Ti | 0.550, 0.500, 0.461 |

Supplementary Table 4. Crystal structure of the monoclinic local structure *Cm* (two-phase model) in PYN-0.4PT. The parameters are refined to fit the experimental PDF in the long range.

|  | *Cm* |
|---|---|
| Lattice Parameters | |
| $a$ (Å) | 5.78004 |
| $b$ (Å) | 5.77474 |
| $c$ (Å) | 4.10920 |
| $\beta$ (°) | 90.359 |
| Atomic positions ($x, y, z$) | |
| Pb1 | 0, 0, 0 |
| O1 | 0.443, 0.000, 0.032 |
| O2 | 0.237, 0.252, 0.466 |
| Yb/Nb/Ti | 0.472, 0.000, 0.542 |

Supplementary Table 5. Crystal structure of the monoclinic local structure *P4mm* (two-phase model) in PYN-0.6PT. The parameters are refined to fit the experimental PDF in the long range.

|  | *P4mm* |
|---|---|
| Lattice Parameters | |
| $a$ (Å) | 4.00374 |
| $c$ (Å) | 4.14237 |
| Atomic positions ($x, y, z$) | |
| Pb1 | 0, 0, 0 |
| O1 | 0.500, 0.000, 0.465 |
| O2 | 0.500, 0.500, 0.092 |
| Yb/Nb/Ti | 0.500, 0.500, 0.547 |

Supplementary Table 6. Crystal structure of the monoclinic local structure *Pm* (two-phase model) in PYN-0.6PT. The parameters are refined to fit the experimental PDF in the long range.

|  | *Pm* |
|---|---|
| Lattice Parameters |  |
| $a$ (Å) | 4.01378 |
| $b$ (Å) | 3.99461 |
| $c$ (Å) | 4.14286 |
| $\beta$ (°) | 90.197 |
| Atomic positions ($x, y, z$) |  |
| Pb1 | 0, 0, 0 |
| O1 | 0.550, 0.000, 0.443 |
| O2 | 0.551, 0.500, 0.060 |
| O3 | 0.042, 0.500, 0.383 |
| Yb/Nb/Ti | 0.539, 0.500, 0.454 |